# Deep Learning on Operational Facility Data Related to Large-Scale Distributed Area Scientific Workflows

Alok Singh, Ilkay Altintas
San Diego Supercomputer Center
University of California, San Diego
La Jolla, CA, USA
{a1singh, ialtintas}@ucsd.edu

Eric Stephan, Malachi Schram
Pacific Northwestern National Laboratory
Richland, WA, USA
{eric.stephan, malachi.schram}@pnnl.gov

*Abstract*—Distributed computing platforms provide a robust mechanism to perform large-scale computations by splitting the task and data among multiple locations, possibly located thousands of miles apart geographically. Although such distribution of resources can lead to benefits, it also comes with its associated problems such as rampant duplication of file transfers increasing congestion, long job completion times, unexpected site crashing, suboptimal data transfer rates, unpredictable reliability in a time range, and suboptimal usage of storage elements. In addition, each sub-system becomes a potential failure node that can trigger system wide disruptions. In this vision paper, we outline our approach to leveraging Deep Learning algorithms to discover solutions to unique problems that arise in a system with computational infrastructure that is spread over a wide area. The presented vision, motivated by a real scientific use case from Belle II experiments, is to develop multilayer neural networks to tackle forecasting, anomaly detection and optimization challenges in a complex and distributed data movement environment. Through this vision based on Deep Learning principles, we aim to achieve reduced congestion events, faster file transfer rates, and enhanced site reliability.

## I. INTRODUCTION

Distributed computing platforms provide a robust mechanism to perform large-scale computations by splitting the task and data among multiple locations, possibly located thousands of miles apart geographically. This distribution of resources can lead to benefits such as redundancy of data and engagement with scientific teams that involve a diverse group of domain experts distributed spatially around the globe. It allows scientists to share their expertise and data with everyone and achieve insights by deploying collective intelligence of entire team. However, this distributed computing comes with its associated problems such as rampant duplication of file transfers increasing congestion, long job completion times, unexpected site crashing, suboptimal data transfer rates, unpredictable reliability in a time range, overshooting congestion, suboptimal usage of storage elements. In addition, each sub-system becomes a potential failure node that can trigger system wide disruptions.

In this vision paper, we outline our plans to leverage Artificial Intelligence methods to tackle operational bottlenecks. Supervised learning methods are considered suitable for tackling complex systems due to their flexibility and efficient training/testing framework. Unsupervised learning techniques lend themselves to interpretation and can tackle uncertainty smoothly and provide mechanisms to infuse domain expertise. Neural networks can be deployed to detect anomalies in dynamic environment with training. Deep Learning algorithms involve development of multilayer neural networks to solve forecasting, classification and clustering solutions. Our approach leverages such Deep Learning algorithms to discover solutions to problems associated with having computational infrastructure that is spread over a wide area.

Our past work in modular learning [1] and large sensor networks [2] focuses on using Machine Learning methods to solve workflow performance issues and leverage large sensor networks to achieve robust cloud performance. In continuation of broader scheme of applying Artificial Intelligence Techniques for designing efficient systems, we present our vision to explore the operational data, extract insights and patterns from it and deploy intelligent systems on Belle II that can provide long-term efficiency gains without much intervention. Developments in Deep Neural Networks in wide areas naturally create scope for their deployment in distributed systems such as Belle II. We believe such efforts will not only shed a light into operational health of experimental facilities and computing infrastructure, but also provide a wealth of information that can be used for dynamic data-driven scheduling of scientific workflows.

In recent years, there have been extensive developments in deployment of neural networks to solve challenging problems [3]–[7]. [8]–[11] and [12] provide a detailed and generic outline of Deep Learning architectures. However, in this paper, we only discuss Deep Learning techniques in the context of improving the Belle II experiment's robustness and efficiency, and develop multilayer neural networks to tackle forecasting, anomaly detection and optimization challenges in a complex and distributed data movement environment. DL techniques can be broadly classified into two areas: *discriminative models* (such as Deep Neural Networks, Recurrent Neural Networks, Convolutional Neural Networks etc.) and *generative models* (such as Deep Boltzmann Machines, Regularized Autoencoders, Deep Belief Networks etc.) [13], [14]. We plan to use these techniques to forecast insights that can improve system-level robustness of Belle II and detect anomalies to improve performance. In conjunction with experimenting with networks such as Convolutional Neural Networks, Recurrent Neural Networks, and Long Short-Term Memory architectures, we will investigate optimal settings for tuning techniques such as regularization, early termination, dropout and dimensionality reduction techniques (e.g., Max Pooling) to achieve best performance.

Deep Learning (DL) has proven surprisingly successful when large amounts of data is available to capture the underlying phenomenon. In the context of Belle II experiments, we capture large amounts of operational metrics characterizing the system performance and data movement across different storage elements. Using Deep Learning Principles, we aim to achieve reduced congestion events, faster file transfer rates, and enhanced site reliability, which in turn will lead to reduced average job completion time. Domain independence of neural networks can allow us to deploy deep networks to forecast robustness of a site dynamically: how likely is a site crash. We will investigate their utility to decipher future behavior such as expected data transfer rates, forecasting congestion and delays, predict re-usability factor for each data transfer, and anticipate overload conditions before they occur in Belle II.

In summary, this forecasting capability can greatly enhance schedulers' performance and improve robustness of the Belle II experiment by providing actionable insights. Belle II experiment's massive data generation provides us an opportunity to leverage Deep Learning algorithms for extracting meaningful patterns and insights from operations logs. As the main contribution of this paper, we have summarized specific algorithms that we plan to deploy in order to perform analytics on Belle II operational datasets, and made an attempt to map each to an expected outcome, that will improve the performance and robustness of the Belle II experiment. In particular, the specific goals of this vision paper are to share our design board with the workflow community and outline our plans to:

- Investigate the effectiveness of Deep Learning Techniques to improve overall efficiency and reliability of experimental facilities managed by large-scale distributed scientific workflows (Section III);

- Examine the performance of Recurrent Neural Networks and Hierarchical Temporal Memory Models for predicting future demand of a given dataset in Belle II experiment, and predict a remote site's availability status dynamically (Section III-C, D);

- Measure the effectiveness of Long Short-Term Memory Models and Autoencoders for carrying anomaly detection by measuring deviation from anticipated (predicted) execution metric or system status (Section III-A, E); and

- Document the outcomes of deploying the Reinforcement Learning approach in making execution of scientific workflows cost and time efficient, using Belle II as a representative use case (Section III-F).

The rest of this paper is organized as follows: Section I starts with an overview of the broader field of research and builds the context and motivation for our current research problem, Section II overviews the representative use case and explains Belle II and ProvEn Framework, Section III presents a mapping of the DL techniques we plan to deploy and corresponding expected benefits we anticipate, and Section IV discusses the overall approach and provides concluding remarks.

## II. EXPERIMENTAL DATA COLLECTION

Deep Learning algorithms mentioned in Section I will allow exploratory analysis of operational data collected from the Belle II experiment. The Belle II experiment itself is a massive collaborative effort involving geographically separated sites and generates large amounts of data, hence this effort to leverage Deep Learning algorithms has immense potential of tackling the problems of scalability and generalization to new scenarios generated due to the size of the experiments. We will use the ProvEn Framework for data collection for analysis. In this section, we summarize Belle II and ProvEn as a background for the rest of this paper.

### A. Belle II

The Belle II experiments [15], [16], designed to probe the interactions of the most fundamental constituents of our universe, is expected to generate about 25 peta bytes (25 x $10^{15}$ bytes) of raw data per year with an expected total storage to reach over 350 peta bytes at the end of the experiment. Data is generated from the physical experiment through the Belle II detector, from Monte Carlo simulations, and user analysis. Data is processed and re-processed through a complex set of operations, which is followed by analysis in a collaborative manner. Users, data, storage and compute resources are geographically distributed across the globe offering a complex data intensive workflow as a case study.

Workflows in Belle II are managed with DIRAC workflow management system [17]. DIRAC is a collaborative tool that provides several key tools for distributed computing, such as data movement, job submissions, and monitoring of jobs. DIRAC uses a central task queue for workload management. The notion of Pilot Jobs is used to manage uncertainties emerging from an inherently unstable resource environment. The DIRAC framework provides a stable workflow management system that has been in used for several years, however, cost and energy efficient use of the grid resources have not been implemented. These features could become important in the context of Belle II operational cost in light of the emerging new hardware architecture and opportunistic resources.

### B. ProvEn Framework

Provenance Environment (ProvEn) is a software platform [18] developed to collect and manage disclosed provenance and performance metrics through data hybridization. ProvEn uses hybridization to synchronize captured provenance such as software process history and data lineage in detailed semantic graph form with environmental factors (e.g., CPU's, storage, networking, I/O, etc.) and timing information to study anomalies and trending patterns occurred at runtime. ProvEn consists of database services and client tools used to collect, store, and provide access provenance information and observed metrics. Figure 1 shows the main ProvEn services to collect provenance data streams, to persist provenance in a semantic store (used for traditional workflow provenance), and to collect streaming metrics (for instance CPU, I/O, and memory activities) in a time series store. ProvEn features a RESTful service layer that is capable of collecting provenance and metrics disclosed by applications using the ProvEn client API, or a harvester capable of extracting provenance from log files using a simple delimited format. As shown in Figure 2, ProvEn's service layer enables any analytical environment, e.g., Jupyter notebooks. In [2], we explained how the framework was used for supervised learning to collect and serve observations used as training data.

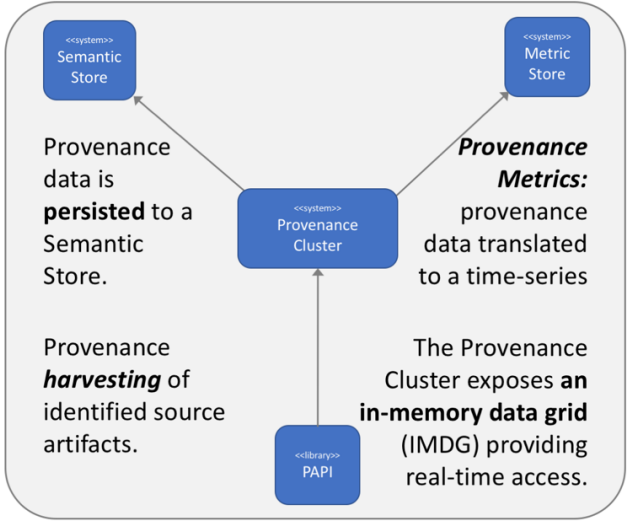

*Figure 1. Provenance Environment (ProvEn) architecture has four key components: a semantic store, a metric store, an in-memory grid and a provenance API.*

III. ANALYSIS OF BELLE II OPERATIONAL DATA FOR INSIGHTS

The abovementioned Belle II workflows currently lack energy efficient use of the distributed computing and data resources. Implementation of coordination mechanisms and workflow schedulers can help reduce the Belle II operational costs in the long term. However, such workflow coordination requires ongoing knowledge of resource status, availibility and load. In this section, we share some of the analytical techniques that can be used to analyze Belle II data with a goal to improve overall efficiency and reliability of Belle II large-scale distributed scientific workflows. Although we pick Belle II as the demonstrating example here, the analytical techniques we discuss in the rest of this section are transferrable to other facilities and experiments in different domains.

*A. Anomaly Detection in Belle II using Autoencoders*

Autoencoders are models which have same size of input and output layers, with a low dimensional middle layer. The target of the autoencoder [19]–[21] is to reconstruct the input by training the neural network, which has a low dimensional layer sandwiched between the input and the output. This forces the model to learn the essential and critical features of input, that are required to rebuild it from this low-dimensional representation. Once an autoencoder is trained on normal data from Belle II operational data and captures its low dimensional features during the training phase, a threshold deviation of reconstruction error can be used to mark anomalous readings. Since, the model performs well on normal data, any reconstruction error can be attributed to anomalous reading. Autoencoders provide a clever way to performing dimensionality reduction and anomaly detection, through the use of identity transformation using a specific constraint. It must be noted that amount of normal data used during training influences model's ability to reconstruct input, hence large amounts of operational data will be required for this experiment. We anticipate that Belle II operations will provide sufficient amounts of training data for building autoencoder based anomaly detectors.

*B. Frequency Domain Analysis in Belle II*

Frequency domain representations of time series operational data can enable deep learning model's to microscopically capture frequency domain patterns and make predictions. In another related problem of speech recognition, which involves a dynamic time series signal, frequency based invariance has been found to be more critical in comparison to temporal invariance [22]–[24]. We would like to investigate Belle II data by transforming time-series logs using Fourier Transforms and measure any changes in Deep Learning Models' ability to make accurate forecasts about future expected output. Frequency domain analysis using Neural Networks can lead to improved understanding of how time-series signals can be tackled using modern deep learning techniques.

*C. Using Convolutional Neural Networks in Belle II*

Convolutional neural networks (CNN or covnets) leverage translational invariance using weights that are shared among different entities. This results from combination of convolutional layer which performs weight sharing and a pooling layer that enables down-sampling between layers. Traditional, CNNs have proved effective for tasks such as computer vision [25]–[27]. Strategic improvements in pooling layer design have proved effective for voice recognition [24], a problem similar to ours i.e. the task of understanding what a time varying signal is communicating. We would investigate the Hierarchical Temporal Memory (HTM) model [28], [29] which extends covnets to include time domain and involve two-way information flow as opposed to uni-directional flow in convolutional neural networks.

*D. Forecasting Intelligent Data Replication Decisions in Belle II using Recurrent Neural Networks*

Data obtained from sensors or logs in a scientific experiment such as Belle II are functions of time, in most scenarios. Recurrent Neural Network (RNN) architecture gives us a structure to handle and learn from a sequence of inputs, such as time-series data. RNNs faced difficulties such as vanishing or exploding gradients [30] in their early stages. Recent developments have tackled gradient issues elegantly and made RNNs reliably robust for training [9], [31]. These developments include use of Hessian free optimizations [32] and improvements of stochastic gradient descent methods [33], [34]. RNNs have ability to keep contextual knowledge of recent or far history while making a forecast as opposed to assuming independence between two different inputs. This ability to progressively learn from a series of inputs give RNNs ability to make contextual predictions for a signal.

We would like to train RNN based models to learn from series of sensor inputs such as file transfer rates over time, job completion durations during months of the year, site-availability as a function of time etc., and measure their ability to make accurate forecasts of events that give us actionable insights. For example, if an RNN can predict site availability in advance, we can use this information to make better scheduling decisions, and avoid problems later. Their success in handling sequential data such as in language modeling [35] and speech recognition [36], [37] bring them under our radar for tackling sequence of sampled data from Belle II sensors.

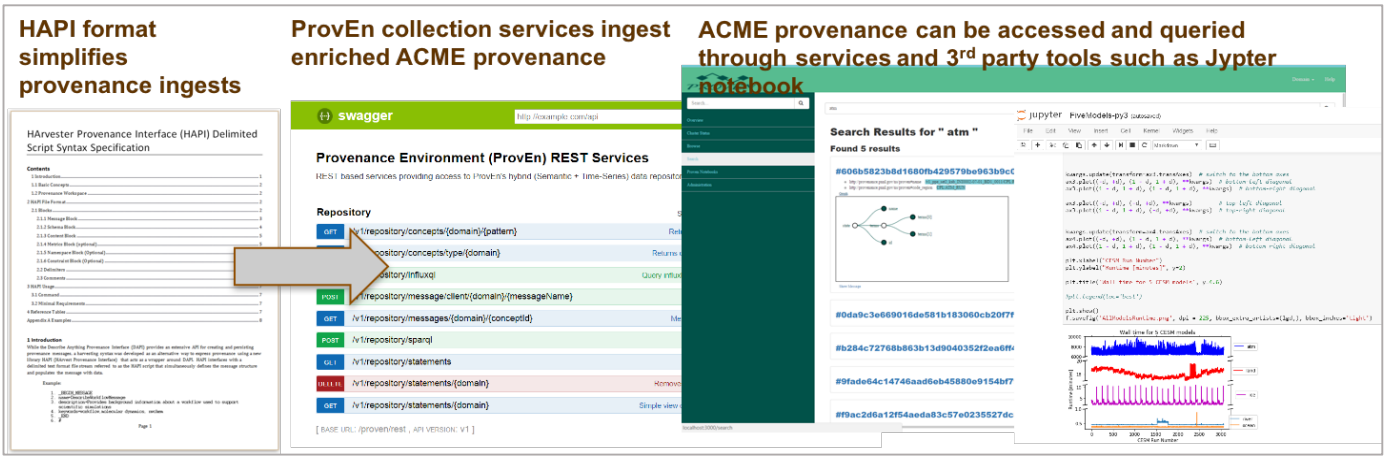

*Figure 2. The Harvester Provenance Interface (HAPI) ingests and enriches information from distributed sources and loads provenance via REST Services. The provenance information is passed to Jupyter notebooks for streamlined analysis such as Pattern Recognition and Machine Learning.*

RNNs have been susceptible to problems such as vanishing or exploding gradients. Hence, we plan to deploy gating techniques such as Long Short-Term Memory and Gated Recurrent Unit to tackle these gradient issues during the training. RNNs have a proven track record of performing well in variable length inputs such as sentiment analysis that attempts to predict whether a movie review is positive or negative, or speech translation that consumes a variable length input and generates a variable length output in a different language.

We anticipate that problems of prediction states of a computation site can be mapped to solving a problem of making a Boolean forecast of a time-series that is composed of 0s and 1s. A 0 at time *t* would signify that the site is not available and 1 would claim otherwise. This temporal output based on a perpetual stream of input from Belle II can generate specialized RNNs that are responsible for forecasting availability for a given site. We will research signals that can results in higher precision.

Another problem to tackle using RNNs is curbing contention due to spike in data transfers at a given time stamp. Maintaining a copy of a dataset at a location, speeds up job completion time for future jobs and reduces chances of congestion. However, we cannot replicate every dataset at every location, due to storage, network and computational constraints. To this end, we need a smart strategy that replicates data at strategic locations in anticipation of future jobs that will need this data. RNNs provide a viable architecture for forecasting future demand of a given dataset. RNNs trained to make prediction about expected usage of a given dataset, can enable intelligent scheduling of data transfers that avoid congestion and also provide data 'on-time' exactly when a new computation may need it. The time-series nature of predicting demand for datasets maps efficiently to RNNs ability to handle a sequence of inputs and generate a sequence of output. Dynamically changing inputs signals will enable RNN model to make new predictions for future time stamps, based on updated information received from sensors or log files. In addition, our research will explore similar deep learning techniques to make smart data replication decisions in anticipation of future demands for data sets.

*E. Long Short Term Memory Networks to Preemptively Detect Anomalies in Belle II*

LSTM [38] networks are specialized recurrent neural networks that manifest concept of short term and long term memory, and make a prediction by combining historical information with newly arrived data. LSTM models have enhanced RNNs in the context of sequence classification [39], [40]. LSTMs have proven to be useful in detecting anomalies in time series data in recent years. LSTMs take into account previous layers input and can simulate forgetful behavior by choosing to partially or completely drop that information. In this sense, we can train LSTMs to make predictions of expected values in future and measure deviations in time series data to capture anomalies.

This is equivalent to training a model on known behavior and classifying any 'other' pattern as anomalous. The advantage of this approach comes from plethora of 'normal' readings available from monitoring Belle II operations. In any large scale system, the amount of 'normal' operational data is disproportionately larger than data about 'malfunctions', as is expected. Hence, we use this normal data set to train a model on what to expect. However, our model becomes susceptible to false positives (classifying normal behavior as abnormal), if our training data does not cover entire range of patterns in time series that pertain to normal behavior. Hence, when we analyze the performance of LSTMs on Belle II operational data in the context of sampled data, we will investigate model's ability to capture anomalies, as a function of training data size, and measure changes in model performance with respect to training data set size and its characteristics.

*F. Reinforcement Learning on Belle II Operational Data*

In recent years, there has been significant improvements in the field of reinforcement learning for a variety of tasks. Reinforcement Learning has potential for drastically improving performance of operations in systems build from a network of computers. [41] provide an entry level explanation of reinforcement learning and its ecosystem. [42] gives a detailed survey of Reinforcement Learning. The central idea of reinforcement learning is to build an agent which observes its environment, decides which action to take, and is receives a reward for the action performed. The reward can be positive for a desired outcome and negative for an undesired outcome. Reinforcement Learning algorithms have shown extraordinary performance under this limited framework which permits only a partial view of the surrounding, and does not depend on prior knowledge of the inner workings of the world it participates in.

In other words, the reinforcement learning agent, learns by trial and error. In computational systems, history of operations can provide much needed data about environment and how it responds to different stimuli. The reward can be characterized by time, space, or event based.

Reinforcement Learning (RL) agents have proved successful in the context of robotics and control [43]–[46]. We aim to build reinforcement learning agents that can learn patterns of Belle II job completion and data transfer, and suggest 'actions' which will result in maximizing rewards. The rewards can be positively or negatively correlated with the goal: a positive reward for each job completed before deadline, a negative reward for performing redundant data transfers of files already move once, a negative reward for reducing rate of data transfer etc.

It is noteworthy that RL algorithms do not need constant reward, but can operate in environments where the reward is delayed or comes at the end of a session. The agents can make decisions that seem to jeopardize outcomes in short duration if they result in large gains over a long period of time. This research will involve studying the Belle II operational data and deciding on input sensor streams to be used for feeding as observation to RL agent, construct a set of actions that the agent can take and design a reward policy that gives positive feedback when desired outcomes are achieved. The RL framework appears to map easily to the Belle II environment and provides possibility of building agents that can provide long term benefits by monitoring key performance signals of the experiment.

## IV. Conclusion

In this vision paper, we summarized our vision for data collection and processing on an experimental facility to extract actionable insights from operational data for use in data-driven scheduling of distributed area workflows. To generate such insights, we focused on investigating the effectiveness of Deep Learning algorithms on real operational dataset generated from Belle II experiment. After careful analysis, we have cherry picked the key Deep Learning algorithms that will allow exploratory analysis of operational data collected using the ProvEn Framework from the Belle II experiment. We will primarily deploy Jupyter based Python Notebooks to conduct this analysis.

The Belle II experiment itself is a massive collaborative effort involving geographically separated sites and generates large amounts of data, hence this effort to leverage Deep Learning algorithms has immense potential of tackling the problems of scalability and generalization to new scenarios generated due to the size of the experiments.

Acknowledgment

This work is supported by DOE DE-SC0012630 for IPPD. The content is solely the responsibility of the authors and does not necessarily represent the official views of the funding agencies.

References

[1] A. Singh, M. Nguyen, S. Purawat, D. Crawl, and I. Altintas, "Modular Resource Centric Learning for Workflow Performance Prediction," in *6th Workshop on Big Data Analytics: Challenges, and Opportunities (BDAC)*.

[2] A. Singh et al., "Leveraging large sensor streams for robust cloud control," in *2016 IEEE International Conference on Big Data (Big Data)*, 2016, pp. 2115–2120.

[3] B. K. Bose, "Expert system, fuzzy logic, and neural network applications in power electronics and motion control," *Proc. IEEE*, vol. 82, no. 8, pp. 1303–1323, Aug. 1994.

[4] I. Kaastra and M. Boyd, "Designing a neural network for forecasting financial and economic time series," *Neurocomputing*, vol. 10, no. 3, pp. 215–236, Apr. 1996.

[5] M. A. Shahin, M. B. Jaksa, and H. R. Maier, "Artificial neural network applications in geotechnical engineering," *Aust. Geomech.*, vol. 36, no. 1, pp. 49–62, 2001.

[6] G. W. Irwin, K. Warwick, and K. J. Hunt, *Neural network applications in control*. IET.

[7] A. S. Miller, B. H. Blott, and T. K. Hames, "Review of neural network applications in medical imaging and signal processing," *Med. Biol. Eng. Comput.*, vol. 30, no. 5, pp. 449–464, Sep. 1992.

[8] Y. Bengio, "Learning Deep Architectures for AI," *Found. Trends® Mach. Learn.*, vol. 2, no. 1, pp. 1–127, Nov. 2009.

[9] Y. Bengio, A. Courville, and P. Vincent, "Representation Learning: A Review and New Perspectives," *IEEE Trans. Pattern Anal. Mach. Intell.*, vol. 35, no. 8, pp. 1798–1828, Aug. 2013.

[10] J. Schmidhuber, "Deep learning in neural networks: An overview," *Neural Netw.*, vol. 61, pp. 85–117, Jan. 2015.

[11] I. Arel, D. C. Rose, and T. P. Karnowski, "Deep Machine Learning - A New Frontier in Artificial Intelligence Research [Research Frontier]," *IEEE Comput. Intell. Mag.*, vol. 5, no. 4, pp. 13–18, Nov. 2010.

[12] L. Deng and D. Yu, "Deep Learning: Methods and Applications," *Found. Trends® Signal Process.*, vol. 7, no. 3–4, pp. 197–387, Jun. 2014.

[13] K. P. Murphy, *Machine Learning: A Probabilistic Perspective*. MIT Press, 2012.

[14] L. Deng and X. Li, "Machine Learning Paradigms for Speech Recognition: An Overview," *IEEE Trans. Audio Speech Lang. Process.*, vol. 21, no. 5, pp. 1060–1089, May 2013.

[15] V. Bansal, M. Schram, and I. Belle Collaboration, "Belle II grid computing: An overview of the distributed data management system.," in *APS Meeting Abstracts*, 2017.

[16] S. Pardi, G. de Nardo, and G. Russo, "Computing at Belle II," *Nucl. Part. Phys. Proc.*, vol. 273, pp. 950–956, Apr. 2016.

[17] S. Paterson, J. Closier, and the L. D. Team, "Performance of combined production and analysis WMS in DIRAC," *J. Phys. Conf. Ser.*, vol. 219, no. 7, p. 072015, 2010.

[18] T. Elsethagen et al., "Data provenance hybridization supporting extreme-scale scientific workflow applications," in *2016 New York Scientific Data Summit (NYSDS)*, 2016, pp. 1–10.

[19] Y. Bengio, P. Lamblin, D. Popovici, and H. Larochelle, "Greedy Layer-wise Training of Deep Networks," in *Proceedings of the 19th International Conference on Neural Information Processing Systems*, Cambridge, MA, USA, 2006, pp. 153–160.

[20] L. Deng, M. Seltzer, D. Yu, A. Acero, A. Mohamed, and G. Hinton, *INTERSPEECH 2010 Binary Coding of Speech Spectrograms Using a Deep Auto-encoder*. .

[21] G. E. Hinton and R. R. Salakhutdinov, "Reducing the dimensionality of data with neural networks," *Science*, vol. 313, no. 5786, pp. 504–507, Jul. 2006.

[22] O. Abdel-Hamid, L. Deng, and D. Yu, "Exploring convolutional neural network structures and optimization techniques for speech recognition.," in *Interspeech*, 2013, pp. 3366–3370.

[23] O. Abdel-Hamid, A. r Mohamed, H. Jiang, and G. Penn, "Applying Convolutional Neural Networks concepts to hybrid NN-HMM model for speech recognition," in *2012 IEEE International Conference on Acoustics, Speech and Signal Processing (ICASSP)*, 2012, pp. 4277–4280.

[24] L. Deng, O. Abdel-Hamid, and D. Yu, "A deep convolutional neural network using heterogeneous pooling for trading acoustic invariance with phonetic confusion," in *2013 IEEE International Conference on Acoustics, Speech and Signal Processing*, 2013, pp. 6669–6673.

[25] D. Ciresan, A. Giusti, L. M. Gambardella, and J. Schmidhuber, "Deep Neural Networks Segment Neuronal Membranes in Electron Microscopy Images," in *Advances in Neural Information Processing Systems 25*, F. Pereira, C. J. C. Burges, L. Bottou, and K. Q. Weinberger, Eds. Curran Associates, Inc., 2012, pp. 2843–2851.

[26] A. Krizhevsky, I. Sutskever, and G. E. Hinton, "ImageNet Classification with Deep Convolutional Neural Networks," in *Advances in Neural Information Processing Systems 25*, F. Pereira, C. J. C. Burges, L. Bottou, and K. Q. Weinberger, Eds. Curran Associates, Inc., 2012, pp. 1097–1105.

[27] M. D. Zeiler, "Hierarchical Convolutional Deep Learning in Computer Vision," Ph.D., New York University, United States -- New York, 2013.

[28] D. George, "How the brain might work: A hierarchical and temporal model for learning and recognition," Ph.D., Stanford University, United States -- California, 2008.

[29] J. Hawkins, S. Ahmad, and D. Dubinsky, "Hierarchical temporal memory including HTM cortical learning algorithms," *Techical Rep. Numenta Inc Palto Alto Httpwww Numenta ComhtmovervieweducationHTMCorticalLearningAlgorithms Pdf*, 2010.

[30] S. Hochreiter, *Untersuchungen zu dynamischen neuronalen Netzen. Diploma thesis, Institut für Informatik, Lehrstuhl Prof. Brauer, Technische Universität München*. 1991.

[31] R. Pascanu, T. Mikolov, and Y. Bengio, "On the difficulty of training recurrent neural networks," in *International Conference on Machine Learning*, 2013, pp. 1310–1318.

[32] J. Martens, "Deep learning via Hessian-free optimization," in *Proceedings of the 27th International Conference on Machine Learning (ICML-10)*, 2010, pp. 735–742.

[33] Y. Bengio, "Deep Learning of Representations: Looking Forward," in *Statistical Language and Speech Processing*, 2013, pp. 1–37.

[34] I. Sutskever, "Training recurrent neural networks," 2013.

[35] T. Mikolov, M. Karafiát, L. Burget, J. Cernockỳ, and S. Khudanpur, "Recurrent neural network based language model.," in *Interspeech*, 2010, vol. 2, p. 3.

[36] G. Mesnil, X. He, L. Deng, and Y. Bengio, "Investigation of recurrent-neural-network architectures and learning methods for spoken language understanding.," in *Interspeech*, 2013, pp. 3771–3775.

[37] K. Yao, G. Zweig, M.-Y. Hwang, Y. Shi, and D. Yu, "Recurrent neural networks for language understanding.," in *Interspeech*, 2013, pp. 2524–2528.

[38] S. Hochreiter and J. Schmidhuber, "Long Short-Term Memory," *Neural Comput.*, vol. 9, no. 8, pp. 1735–1780, Nov. 1997.

[39] A. Graves, N. Jaitly, and A. Mohamed, "Hybrid speech recognition with Deep Bidirectional LSTM," in *ASRU*, 2013.

[40] A. Graves, "Sequence transduction with recurrent neural networks," *ArXiv Prepr. ArXiv12113711*, 2012.

[41] R. S. Sutton and A. G. Barto, *Introduction to Reinforcement Learning*, 1st ed. Cambridge, MA, USA: MIT Press, 1998.

[42] L. P. Kaelbling, M. L. Littman, and A. W. Moore, "Reinforcement learning: A survey," *J. Artif. Intell. Res.*, vol. 4, pp. 237–285, 1996.

[43] S. Schaal and C. G. Atkeson, "Robot juggling: implementation of memory-based learning," *IEEE Control Syst.*, vol. 14, no. 1, pp. 57–71, 1994.

[44] S. Mahadevan and J. Connell, "Automatic programming of behavior-based robots using reinforcement learning," *Artif. Intell.*, vol. 55, no. 2–3, pp. 311–365, 1992.

[45] M. J. Mataric, "Reward functions for accelerated learning," in *Machine Learning: Proceedings of the Eleventh international conference*, 1994, pp. 181–189.

[46] R. H. Crites and A. G. Barto, "Improving elevator performance using reinforcement learning," in *Advances in neural information processing systems*, 1996, pp. 1017–1023.